# Enhanced Instantaneous Elastography in Tissues and Hard Materials Using Bulk Modulus and Density Determined without Externally Applied Material Deformation

Yuqi Jin, Ezekiel Walker, Arkadii Krokhin, Hyeonu Heo, Tae-Youl Choi and Arup Neogi*

*Abstract*—Ultrasound is a continually developing technology that is broadly used for fast, non-destructive mechanical property detection of hard and soft materials in applications ranging from manufacturing to biomedical. In this study, a novel monostatic longitudinal ultrasonic pulsing elastography imaging method is introduced. Existing elastography methods require an acoustic radiational or dynamic compressive externally applied force to determine the effective bulk modulus or density. This new, passive M-mode imaging technique does not require an external stress, and can be effectively utilized for both soft and hard materials. Strain map imaging and shear wave elastography are two current categories of M-mode imaging that show both relative and absolute elasticity information. The new technique is applied to hard materials and soft material tissue phantoms for demonstrating effective bulk modulus and effective density mapping. As compared to standard techniques, the effective parameters fall within 10% of standard characterization methods for both hard and soft materials. As neither the standard A-mode imaging technique nor the presented technique require an external applied force, the techniques are applied to composite heterostructures and the findings presented for comparison. The presented passive M-mode technique is found to have enhanced resolution over standard A-mode modalities.

*Index Terms*—Ultrasonic imaging, bulk modulus elastography, density mapping, longitudinal pulse, tissue phantoms contrast.

## I. Introduction

Ultrasound technologies are continuously developing and enjoy broad usage in biomedical and manufacturing engineering applications for sensors [1], cleaning [2], welding [3], material characterization [4], and imaging [5]. In manufacturing engineering, ultrasonic characterization is an alternative test method for mechanical properties that is faster than tensile and compression tests on stiff, isotropic materials such as metals and alloys [6]. By utilizing both the measured longitudinal and transverse speed of sound, the sample thickness, and material density, mechanical properties such as the Young's modulus, shear modulus, and Poisson's ratio can be extrapolated [4]. However, many methods are commonly insufficient for simultaneous application to soft materials like biological tissues due to the effects of dispersion and attenuation [7].

Ultrasonic imaging, which is widely used across many disciplines in both academic and industrial settings, incorporates ultrasonic characterization techniques for visualization. Basic ultrasonic imaging (B mode imaging) is based on time-of-flight measurement of the ultrasound pulse, and is often displayed as a greyscale map where intensities are connected to an elastic property [8]. Conversely, A-mode imaging measures the energy level (amplitude) of reflected waves at a fixed distance [9]. Resolution of the imaging modes are closely related to wave frequency for the axial direction, and beam waist size for the lateral direction. Frequencies between 1 to 20 MHz are most often used [5]. Higher frequency devices are used to improve imaging resolution, with frequencies reaching as high as 100 MHz [10] or even hypersound [11]. For ultrasonic attenuation, measures of the ratio of the attenuation coefficient between a sample phantom and reference phantom are collected and mapped in color grades to provide in depth detail of the features of a sample [12].

Ultrasound elastography is more commonly used in biomedical applications, with elastographic imaging (M-mode imaging) a particular focus of recent research. Strain map elastography, one of the earliest developed methods in M-mode imaging, is dependent on compressive changes in a composite sample's thickness. Compressive stress on materials with differing elastic properties causes varying degrees of deformation in the linear elastic range. Strain mapping utilizes the change in time of flight information between a sample with and without stress to reconstruct the change in thickness of composites within the sample. Subsequently, the Young's modulus and Poisson ratio are derived [13].

The earliest designs of strain elastography applied compressive force on a sample manually using an ultrasonic

This work is supported by an Emerging Frontiers in Research and Innovation grant from the National Science Foundation (Grant No. 1741677). AN also acknowledges the support from AMMPI 2019 SEED grants.

Yuqi Jin is graduate student now in the Mechanical and Energy Engineering Department, University of North Texas, Denton, TX, 76209 USA. (e-mail: yuqijin@my.unt.edu).

Ezekiel Walker is co-founder and researcher now in Echnovus Inc, Denton, Texas 76205, USA. (e-mail: ezekiel.walker@echonovus.com).

Arkadii Krokhin is professor now with Department of Physics, University of North Texas, Denton, TX 76203 USA. (e-mail: arkady@unt.edu).

Hyeonu Heo is adjunct professor now with the Mechanical and Energy Engineering Department, University of North Texas, Denton, TX, 76209 USA. (e-mail: Hyeonu.Heo@unt.edu).

Tae-Youl Choi is associate professor now with the Mechanical and Energy Engineering Department, University of North Texas, Denton, TX, 76209 USA. (e-mail: tae-youl.choi@unt.edu).

Arup Neogi is distinguished research professor now with Department of Physics, University of North Texas, Denton, TX 76203 USA. (e-mail: arup.neogi@unt.edu).

transducer [14], but further developed to be an automated applied and held force [15]. In practicality, ultrasound elastography is ineffective when a target material is hard, deep, and/or has fluid in interstitial spacings [16]. Additionally, though this method doesn't offer quantitative values for the bulk and Young's modulus, it can distinguish materials which have distinct Poisson's ratios in real time [17].

Techniques that use acoustic radiational force instead of mechanical forces to determine elastic properties fall under impulse strain imaging [18]. From multiple measurements of displacement information with acoustic radiational force, small deformation differences can be found and used for calculating the Young's modulus. Elasticity information is usually represented on a color scale with hard, or stiff, features represented by warm tones and soft, more malleable features by cold tones [19].

Impulse strain mapping and strain map elastography are commonly used methods in both laboratory and commercial settings. Poisson's ratio mapping, however, is a technique looking for vertical strain information similar to strain and impulse strain imaging, but is restricted to laboratory use. For the method, the sample is in water ambient, and a vertical compressional force on the sample raises the water level in the tank. The measured horizontal elongation provides effective Poisson's ratio map calculated on a scale from 0 to 0.5 [20]. In commercial devices, strain elastography or impulse strain mapping are usually used, with either the absolute or relative elastic values represented in a color scale overlapped on grey scale B-mode images [21].

Another popular M-mode imaging modality is shear wave elasticity imaging (SWEI) and is comprised of three methods: Point Shear Wave (PSW) imaging, Surface Shear Wave (SSW) imaging, and Transient Shear Wave (TSW) imaging. Whereas strain elastography or impulse strain mapping primarily are determined with longitudinal waves, SWEI techniques require shear waves. SSW imaging, a relatively recent SWEI method, measures shear wave dispersion and velocity in the temporal domain. The method uses a bistatic setup to provide one focal surface by two overlapped focal zones and combines the overlapped focal zones with an external radiation force to obtain the non-quantitative shear elasticity distribution [22]. PSW and TSW differ from SSW in that they require a monostatic setup.

Point shear wave imaging utilizes shear waves and is used to find either the Young's or shear modulus (elasticity) by measuring the change of the shear wave propagation speed in a focal zone with an applied radiational compressive force [23], lateral force [24], or shear force [25]. SWEI is dependent on adequate sample deformation to modify the measured shear wave speed of sound [26]. For point shear applications, SWEI provides accurate results when used in samples greater than 20mm thick [27] or harder tissue materials such as muscle [28].

In the lower frequency range, Supersonic Shear imaging (SSI), also called Transient Shear imaging, is setup using a monostatic ultrasound probe connected to an external acoustic radiational force generator. The probe continuously records B mode images to find deformation in a specified temporal range after the shear force propagates in the sample [29]. The deformation information forms the basis of a quantitative Young's modulus map.

Shear wave velocity and temporal dispersion, represented quantitatively by echo phase shift due to an external dynamic force, is also used for transient shear imaging in inhomogeneous tissues using a monostatic arrangement [30]. Drawbacks of SWEI in biomedical applications are primarily functions of the impact of body fluids on elastographic results due to the inability of fluid to transmit shear waves [31].

Techniques to image the elastic properties of materials have been primarily geared towards soft, biological samples for biomedical applications, whereas for hard materials the techniques have been used primarily in non-imaging modalities. Effective stiffness (bulk modulus) of monoatomic metal and alloys undergoes significant cavitation erosion due to environmental impact [32]. However, the limitations of current techniques include the need for an external pressure, difficulty in detecting deformation in stiff materials, and the need for a bistatic setup for some methods.

In this study, Effective Bulk Modulus Elastography (EBME) imaging is presented as a new imaging technique. From measuring the acoustic impedance of the scanned sample, effective bulk modulus and effective density mapping can be constructed using classical speed of sound theories $K = \rho c^2 = Zc$ and $\rho = Zc^{-1}$. The method can provide effective stiffness information with bulk modulus scale elastography using a monostatic setup with longitudinal pulses absent any external radiational or cyclidic stress application on the sample. Derivation of the method is given below.

Three experiments were performed to evaluate the efficiency of this new method and are discussed below. In Experiment 1, a hard and soft material combined as a single sample were examined. The hard, low dispersion material was an aluminum slab with a well-defined, large, rectangular area filled with the soft, dispersive material, silicone rubber. For Experiment 2, a series of hard materials in parallel were evaluated using the EBME technique to examine the effectiveness of EBME for distinguishing hard materials monostatically without an external applied stress or direct physical contact. The series of hard materials consisted of copper, PVC plastic, and aluminum. Experiment 3 pertained to application of EBME to soft, tissue-like materials and the capability of EBME to distinguish between soft materials that mimic healthy tissues and calcified or hardened tissues that may indicate ailments. The experiment contrasts three tissue phantoms synthesized using standard formulations for healthy and tumor-like tissue where elastic stiffness values are similar [33, 34, 35].

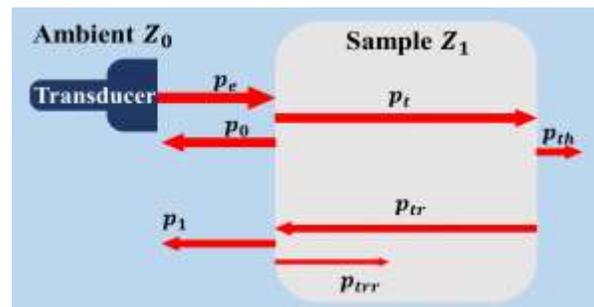

Fig. 1. Acoustic pressure distribution from transducer emitted pulse to the second echo which passed into sample then reflected. $t$ is transmission coefficient and $r$ reflection coefficient. Illustration not to scale.

## II. ANALYTICAL MODEL

In ideal (inviscid) fluid the field of velocities is potential, i.e. $\vec{V} = \nabla \varphi$. In a homogeneous fluid, the scalar potential $\varphi$ satisfies a standard wave equation $c^{-2}\ddot{\varphi} + \nabla^2 \varphi = 0$, where $c$ is speed of sound. The oscillating pressure produced by sound wave is also given by the potential, $p = -\rho \dot{\varphi}$, where $\rho$ is the density of fluid. The proposed mechanism of scanning is based on analysis of short acoustic pulses reflected from elastic samples imbedded into a fluid with known mechanical parameters. Since we operate with short pulses, the amplitude of the reflected signal is not the same as the one obtained for a continuous wave.

Let us consider a reflection of a short pulse $\varphi_e(T)$ emitted by a transducer from a slab shown in Fig. 1. Two output pulses are registered. The first one is the pulse reflected from the left boundary of the slab, $\varphi_0(T)$. The second one coming with a time delay is reflected from the right boundary, $\varphi_{out}(T)$. The duration of the original pulse $\varphi_e(T)$ is so short that it is over by the moment the second pulse hits the left boundary. Due to this condition the left boundary is stationary when the second pulse reaches it. On the other hand, the duration of the input pulse is long as compared to the period of oscillations $2\pi/\omega$ generated by the transducer. The duration of the pulse is also long enough to neglect frequency dispersion of the impedances $Z_0$ and $Z_1$. The latter condition means that the linear relation between the input and output signals can be obtained by replacing the pulses by monochromatic plane waves with angular frequency $\omega$ and corresponding wave vectors $k_0 = \omega/c_0$ and $k_1 = \omega/c_1$.

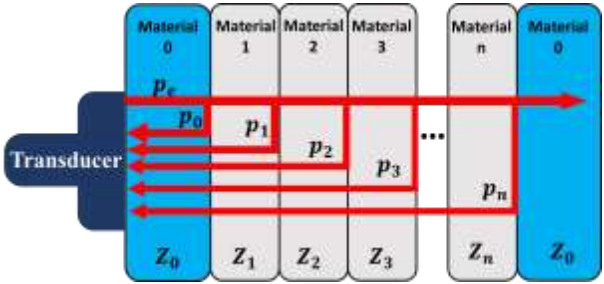

Fig. 2. A layered sample scanned by a series of pulses emitted by a transducer. The first and last layers are of the same ambient material which has a known acoustic impedance $Z_0$. The internal scanned layers are characterized by acoustic impedances $Z_1, Z_2, Z_3, \ldots Z_n$.

The relation between the input signal $\varphi_e$, the first reflected $\varphi_0$ and the transmitted signal $\varphi_t$ is obtained from the boundary conditions at the interface between media 0 and 1, where the acoustic pressures and velocities are continuous,

$$p_e + p_0 = p_t, \qquad (1)$$

$$V_e + V_0 = V_t. \qquad (2)$$

Representing the potentials in the form of plane waves

$$\varphi_e(x,T) = e^{ik_1 x - i\omega t},$$
$$\varphi_0(x,T) = r_{0,1} e^{-ik_0 x - i\omega t}, \qquad (3)$$
$$\varphi_t(x,T) = t_{0,1} e^{ik_1 x - i\omega t}$$

the following known results are obtained from Eqs. (1) and (2) for the reflection and transmission coefficients:

$$t_{0,1} = 2Z_1/(Z_0 + Z_1), \qquad (4)$$

$$r_{0,1} = \frac{Z_1 - Z_0}{Z_1 + Z_0}. \qquad (5)$$

Here indices 0 and 1 are related to the ambient material (water) and to the sample, respectively.

Now the linear relation (2) for velocities can be rewritten through pressures,

$$p_t = \frac{Z_1}{Z_0}(p_e - |p_0|), Z_1 > Z_0, \qquad (6)$$

$$p_t = \frac{Z_1}{Z_0}(p_e + |p_0|), \ Z_1 < Z_0. \qquad (7)$$

As it follows from Eq. (5) the signal $p_0$ is reflected with inverted phase, if $Z_1 < Z_0$. In the experiment only the amplitude $|p_0|$ of the reflected signal is measured, therefore Eqs. (6) and (7) explicitly take into account the phase of the reflected signal.

The unknown impedance $Z_1$ can be calculated if the second reflected signal $p_1$ is measured. This signal originates from $p_t$ which reaches the right boundary of the sample in Fig. 2, and reflects from it. The reflected signal is $r_{1,0} p_t = -r_{0,1} p_t$. The negative sign appears because Eq. (5) is antisymmetric with respect to transformation $Z_1 \leftrightarrow Z_0$. Transmission of the reflected signal through the left boundary of the sample (which remains stationary) reduces its amplitude to

$$p_1 = t_{1,0} r_{1,0} p_t = \frac{2Z_0|Z_1 - Z_0|}{(Z_1 + Z_0)^2}[p_e - \text{sign}(Z_1 - Z_0)|p_0|]. (8)$$

This is the second echo signal received by the transducer. Note that $p_1 > 0$. All three experimentally measured signals, $p_e$, $|p_0|$, and $p_1$ enter to Eq. (8) in a combination $\alpha = \frac{p_1}{p_e - \text{sign}(Z_1 - Z_0)|p_0|}$. If the value of $\alpha$ is known from the experiment, the impedance $Z_1$ can be calculated from quadratic equation (8). Depending on the value of the ratio $Z_1/Z_0$ the solution of this equation can be written as follows:

$$\frac{Z_1}{Z_0} = \frac{-1 - \alpha - \sqrt{4\alpha + 1}}{\alpha - 2}, \qquad \frac{Z_1}{Z_0} > 1, \quad (9)$$

$$\frac{Z_1}{Z_0} = \frac{(1-\alpha) + \sqrt{1-4\alpha}}{\alpha + 2}, \ \frac{1}{3} < \frac{Z_1}{Z_0} < 1, \quad (10)$$

$$\frac{Z_1}{Z_0} = \frac{(1-\alpha) - \sqrt{1-4\alpha}}{\alpha + 2}, \ 0 < \frac{Z_1}{Z_0} \leq \frac{1}{3}. \quad (11)$$

The dependence $Z_1(\alpha)/Z_0$ is plot in Fig. 3 for $0 \leq \alpha \leq 1/4$. The values of the elastic bulk modulus and mass density are calculated from the known impedance.

If the sample consists of $n$ layers of different elastic materials, as shown in Fig. 2, the incoming signal $p_e$ scans all the layers and the transmitter registers $n+1$ echo signals, $|p_0|$, $p_1, p_2, \ldots p_n$. Generalizing Eq. (8) to a multilayer stack, one obtains,

$$p_k = \frac{Z_1}{Z_0}(p_e - \mathrm{sign}(Z_1 - Z_0)|p_0|) \cdot \left(\prod_{i=2}^{k} t_{i-1,i}\right) r_{k-1,k} \left(\prod_{i=1}^{k} t_{i,i-1}\right), k = 1,2,\cdots,n. \quad (12)$$

Here $t_{i-1,i} = (2Z_i)/(Z_{i-1} + Z_i)$ and $r_{k-1,k} = (Z_k - Z_{k-1})/(Z_k + Z_{k-1})$ are the transmission and reflection coefficients, respectfully, for the boundary between the $(n-1)$th and $n$th layers. Since there is the same ambient medium (water) on the both sides of the sample $r_{n-1,0} = (Z_0 - Z_{n-1})/(Z_{n-1} + Z_0)$.

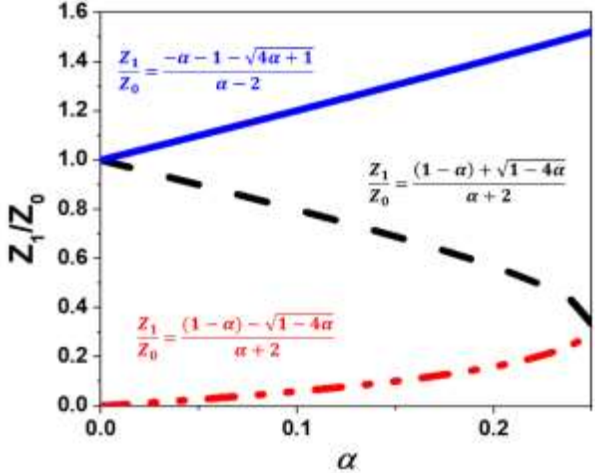

Fig. 3. Multivalue function $Z_1(\alpha)/Z_0$ given by Eqs. (9)-(11).

The recorded echo-pulses are time-dependent functions. From the equipment datasheet, calibrated probe sensitivity was a single coefficient in term of $p/V$, where the $V$ was the signal amplitude in volts. For the numerical values of $|p_0|, p_1, p_2, \ldots p_n$ are taken the corresponding absolute peak values of the measured echo-pulses. Before executing the raster scan the pulse $p_e$ was emitted in the ambient medium without a sample using bistatic calibration. The distance between two transducers is selected to be twice the distance between the emitting transducer's surface and the interface between the ambient material 0 and the first layer in following experiments.

In the case of an $n$-layered sample the measured quantities are

$$\alpha_1 = \frac{p_1}{p_e - \mathrm{sign}(Z_1 - Z_0)|p_0|},$$

$$\alpha_2 = \frac{p_2}{p_e - \mathrm{sign}(Z_1 - Z_0)|p_0|},$$

$$\vdots$$

$$\alpha_n = \frac{p_n}{p_e - \mathrm{sign}(Z_1 - Z_0)|p_0|}. \quad (13)$$

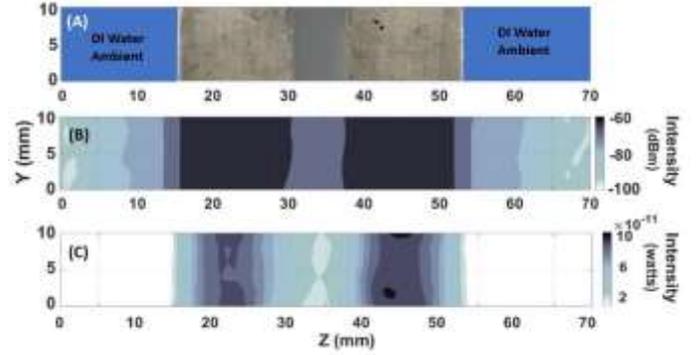

Fig. 4. A-Mode scanned imaging. (A) is whole raster scanned area for visible comparison with (B) and (C). The sample width is 38mm with the silicone rubber filled defect 8mm wide in the center. The rest of the scanned area was DI water ambient. (B) is the traditional A-Mode imaging logarithmically scaled. (C) is traditional A-Mode imaging in linear scale.

If all $\alpha$'s are recorded, the set of $n$ nonlinear equations (12) for the unknown impedances $Z_1, Z_2, \ldots Z_n$ can be solved numerically. We applied the iterative method which requires the initial values for each unknown impedance. These initial values should be selected from reasonable physical considerations based on cursory knowledge of the initial material in which the initial pressure wave is emitted ($Z_0$) and the subsequent material in comparison to the initial material ($Z_1$). If the general ratio of the impedance of the primary to secondary material is known, the solutions for Eq. 12 converge.

The exact position of each reflected pulse within the recorded temporal window, are defined by two instants, $T_{i,k}$ and $T_{f,k}$, corresponding to the beginning and the end of the $k^{\text{th}}$ pulse, i.e. the pulse reflected from the layer with impedance $Z_k$. In the time domain, $T_{i,k}$ and $T_{f,k}$ were found algorithmically.

For this work, the beginning of a pulse is particularly important as it is necessary for determining time of flight delay and subsequently speed of sound. A noise reference level is set by taking the absolute value of the maximum signal amplitude in the first 2% of sampled data points from the calibration of $p_e$. During calibration of $p_e$, the pulse window was set such that the center of $p_e$ was near the center of the time window of collected data points. This procedure greatly reduced the chance of overlap between the calibration pulse and the beginning of the time window used for setting the noise level.

The beginning of a pulse envelope, $T_{i,n}$, is set as the first point in which a continuous $0.05\ \mu s$ of signal, 5% of the pulse width in water, exceeds 105% of the noise reference level. The ensuing end of that pulse envelope, $T_{f,n}$, was calculated by examining the transient data from $T_{i,n}$ and set as the point when a continuous $0.05\ \mu s$ of signal reduced to less than 90% of the noise reference level.

The speed of longitudinal sound $c$ in each layer can be calculated if the time delay between two echo signals obtained as reflections from its boundaries is measured. From impedance and speed of sound the bulk modulus and mass density of each elastic layer are easily calculated as

$$K = \rho c^2 = Zc, \quad (14)$$

$$\rho = \frac{Z}{c}. \quad (15)$$

## III. RESULTS AND DISCUSSION

### A. Experiment 1- Hard and soft material composite sample

Experiment 1 concerns the application of the EBME technique to a sample comprised of both a hard material and a soft material. The scanned sample was a 38mm wide aluminum slab with an 8mm wide rectangular defect filled with silicone rubber as shown in Fig. 4A. Data was collected from a 2D, 70 x 10 mm (lateral x vertical) raster scan in deionized (DI) water ambient. Fig. 4B and Fig 4C represent the intensity data in the logarithmic (4B) and linear (4C) scales. As Fig. 4 (B) shows, logarithmic intensity mapping (A-Mode imaging) clearly shows the silicone rubber filled rectangular defect between two sides of aluminum. The average width of the silicone rubber is around 9mm and average aluminum width of each sides are both around 13mm in the imaging.

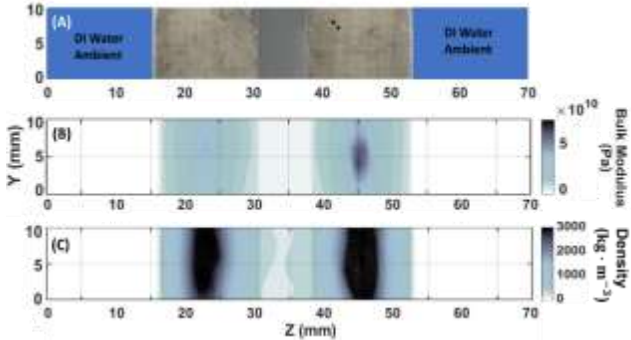

Fig. 5. Effective bulk modulus elastography and effective density mapping of aluminum with silicone on a linear scale. (A) is the raster scanned area for visible comparison with (B) and (C). (B) is the EBME mapping of the bulk modulus (K) from Eq. 14. The color bar range is 0 GPa to 72.5 GPa. (C) is the effective density mapping from Eq. 15.

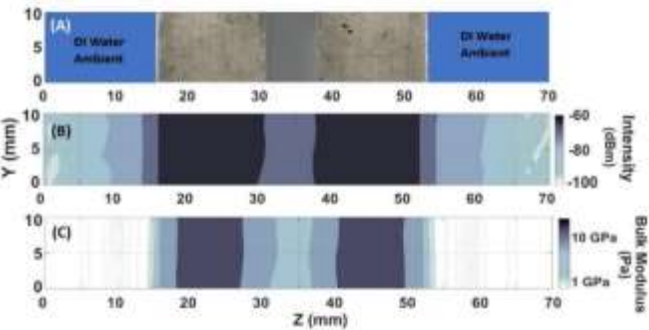

Fig. 6. A-Mode scanned imaging and bulk modulus elastography both logarithmically scaled. (A) Scanned sample area for visible comparison with (B) and (C). (B) Logarithmic scale intensity as shown in Fig. 6B. (C) EBME bulk modulus logarithmically scaled ($log_{10} K$). Compared with linear mapping, the logirithmic scale mapping exhibits visually higher spatial resolution.

Due to the size of transducer and the lateral beam width of the unfocused transducer, large diffraction patterns appear close to both sides of the aluminum. Water ambient exhibits similar contrast to the silicon as is to be expected due to the high impedance mismatch with aluminum, but relatively low impedance mismatch with silicon (Fig. 5B).

Linear scale A-Mode imaging is rarely utilized in practice.

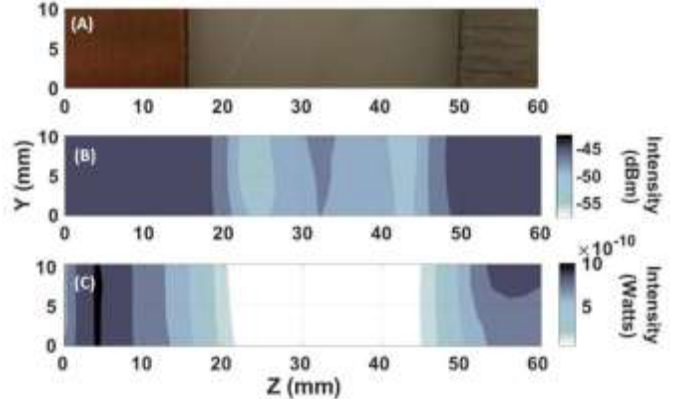

Fig. 7. A-Mode scanned image of hard material composite. (A) is entire raster scanned area for visible comparison with (B) and (C). From left to right, copper is 15mm wide. PVC plastic is 35mm wide. Aluminum is 10mm. (B) is traditional A-mode imaging in logarithmic (dBm) scale. (C) is traditional A-mode imaging in linear (watts) scale.

However, in the linear scale, resolution of the object decreases whereas the contrast increases greatly (Fig. 4C). The intensity gradients in the linear scale, resultant, again, from the transducer size and lateral beam width, make determination of the material boundaries tenuous at best. In comparison with Fig. 4B, the contrast between water, aluminum, and silicone increases greatly which has the potential to aid in the identification of heterostructures. The average width of the scanned sample is around 39mm in linear scale, within 3% of the actual width.

As part of EBME, the bulk modulus and density can be extrapolated from acquired data using Eq. 9 and Eq. 10. Since the method is functionally dependent on the impedance of the ambient medium, and the accurate characterization of the pressure-voltage sensitivity of the detector, the bulk modulus and density are more correctly termed as effective bulk modulus and effective density. Quantitatively correct bulk modulus and density from EBME can be derived when the ambient medium and equipment characterization are accounted for accurately.

For Experiment 1, the effective bulk modulus and density are given in Figs. 5B and 5C. As both the bulk modulus and density are on a linear scale, the impact of the low lateral resolution of the setup becomes apparent. The material boundaries are not well defined due to the gradient values at the transitions between materials. However, as with the intensity, the contrast in the elastic properties between the materials is very clearly seen in the contrasting values, especially at the center of a particular material.

In this setup, the bulk modulus and density values of aluminum are found to be 63.4 GPa and 2720 $kg \cdot m^{-3}$, a 6.76% and 1.47% error respectively as compared with standard techniques [36, 37]. Silicon, a soft material, was found by EBME to have a bulk modulus of 1.9 GPa and density of 1380 $kg \cdot m^{-3}$. Errors as compared to standard methods are -9.7% for the bulk modulus and -9.6% for the density. The relatively close density and bulk modulus values of silicone to water caused a decrease in the signal to noise ratio (SNR) during the data acquisition phase. The silicon rubber had less width than the half size of the transducer surface. The peak values were considered as suitable bulk modulus and density values. As the transition zone showed, the regions besides the location of peak values were scanned partially on the aluminum and the other materials (water or silicone rubber). For this work, to maintain uniformity and focus on the application of the EBME technique, no additional advanced signal processing techniques were performed to increase the SNR for silicon or other soft materials.

The logarithmically scaled EBME is given in Fig. 6, where the bulk modulus (K) is scaled as $\log_{10} K$. The visual impact of the gradients in effective bulk modulus are reduced by logarithmic binning. As compared to the normal A-Mode methods that map intensity, the EBME bulk modulus spatial resolution does not replicate the distinct aluminum-silicone boundaries. However, logarithmic EBME bulk modulus much more clearly delineates between different materials as the silicone in the center of the aluminum is clearly distinguishable from the ambient water. For diagnostics of hard and soft systems, this can be invaluable in determining potential defects that are undesirable versus those that are inconsequential.

## B. Experiment 2 - Hard material composite sample

In the second experiment, a hard material composite consisting of independent blocks of copper, PVC plastic, and aluminum was scanned using the same setup as Experiment 1. The sample was used to evaluate the capability of EBME to distinguish between hard materials. The total scanned area was 10 mm along the vertical, y-axis at 2 mm intervals, and 60 mm along the lateral, z-axis at 1 mm intervals. The widths of each material in the sample area were non-uniform, with copper occupying 15mm of the total width, and PVC and aluminum at 35 mm and 10 mm, respectively.

Fig. 7 gives both the linear and logarithmic A-Mode resultant image from the hard sample composite. The setup is not optimized for high resolution of material bounderies. However, whereas the logarithmic scale does give a relatively accurate representation of the hard material boundaries, the intensity scale alone does not adequately distinguish between copper and aluminum on the far left and far right of Fig. 7B. The linear A-Mode method performs much worse as the boundaries are not clear, and the aluminum and copper are not distinguishable. In both cases, A-mode imaging does show the PVC material as significantly different than both aluminum and copper.

From the EBME derived bulk modulus and density, three distinct materials and their boundaries can be readily recognized. The large contrast in elastic properties between the PVC plastic and metals allows for the estimation of its width from the linear EBME. Using the bulk modulus in Fig. 8B, PVC is 31mm with a bulk modulus of 9.26 GPa resulting in -11% error in the width resolution and 5.5% error from a bulk modulus of 8.75 GPa determined using accepted standard techniques. For copper, the averaged bulk modulus is 138 GPa, about 4.4% off of the value derived from standard techniques. Aluminum has the highest degree of error in the bulk modulus at -10.5%, an averaged 60.7 GPa from EBME versus 68.6 GPa standardized. Density is also relatively well characterized as the EBME derived values from averaging each of the areas of the sample materials comes to 7720 $kg \cdot m^{-3}$, 1480 $kg \cdot m^{-3}$, and 2544 $kg \cdot m^{-3}$ for copper, PVC, and aluminum respectively. Errors for the density as compared with non-EBME techniques are -3.3%, 5.3%, and -5.8%. Additionally, as Fig. 9 shows, logarithmic scaled EBME (Fig. 9C) more accurately represents the hard material distribution shown in Fig. 9A as compared with the logarithmic A-Mode result shown in Fig. 9B.

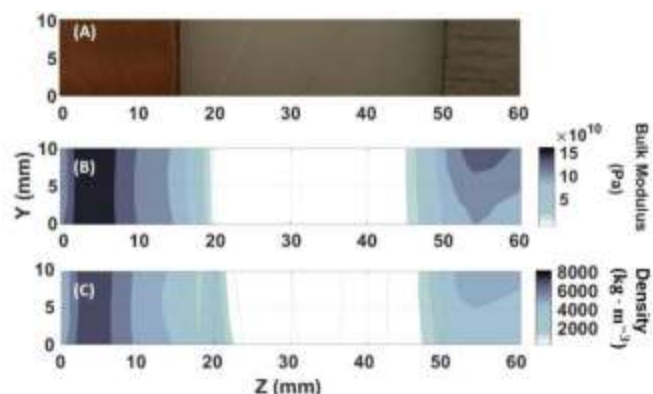

Fig. 8. Effective bulk modulus and effective density mapping linearly scaled. (A) is the raster scanned area for visible comparison with (B) and (C). (B) is EBME from Eq. 14. The color bar is ranged 1 GPa to 160 GPa. (C) is the effective density derived from Eq. 15. The density color bar is ranged 800 $kg \cdot m^{-3}$ to 8000 $kg \cdot m^{-3}$.

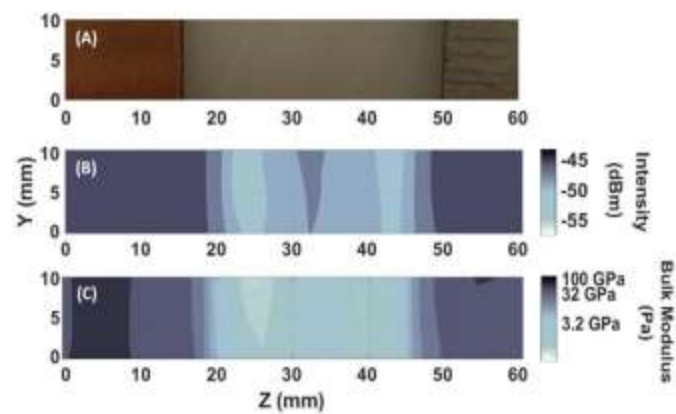

Fig. 9. A-mode scanned imaging and bulk modulus elastagraphy logarithmically scaled. (A) is the raster scanned area for visible comparison with (B) and (C). (B) is A-mode scanned imaging logarithmically scaled. (C) is the bulk modulus using EBME, logarithmically scaled ($log_{10} K$). The color bar is ranged 8.3 to 11.3. 3.2 GPa, 32 GPa, and 100 GPa labels are located on the logarithmic scale color bar.

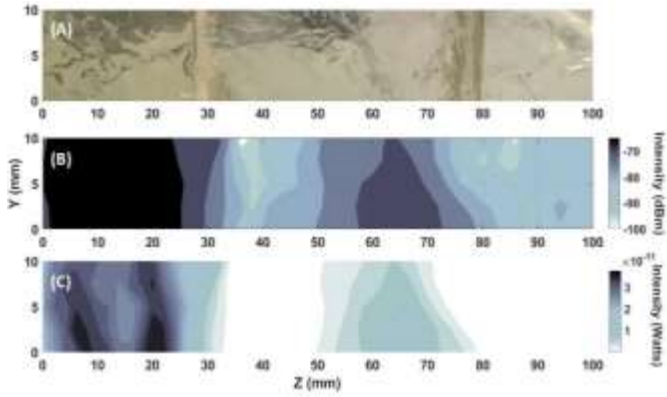

Fig. 10. A-Mode scanning method for Sample 1, comprised of three tissue phantoms with 22.5% (far left), 18% (middle), and 3.5% (far right) gelatin by weight. (A) is whole raster scanned area for visible comparison. (B) Logarithmic scaled intensity. (C) Linear scaled intensity. Only the highest weight % gelatin is clearly distinguishable.

The width from EBME showed about 5mm overestimation on the copper and aluminum side comparing with photograph. The overestimation in width was due to the comparable size of the transducer and the samples. The transducer was 25.4mm (1 inch) diameter large, but the copper block was about 20mm wide and the aluminum was only 10mm wide.

### C. Experiment 3 – Soft tissue phantom composite sample 2

Soft materials, specifically soft materials that mimic organic tissues, present special challenges for ultrasonic characterization. The materials are commonly dispersive and attenuate sound much faster than hard materials. Additionally, tissue-like materials may have features similar to water, making them indistinguishable in standard A-Mode imaging modalities. For Experiment 3, examination of two separate samples comprised of composites of gelatin tissue phantoms was carried out. Sample 1 consisted of gelatin tissue phantoms synthesized using [35], where three (3) gelatin blocks, 22.5%, 18.0%, and 3.5% gelatin respectively, were placed adjacent to each other.

The total scanned area for Sample 1 was 100 x 10 mm, where the 22.5% gelatin tissue phantom was 28mm wide, the 18.0% gelatin tissue phantom 52mm in width, and 3.5% gelatin a width of 20mm. The A-Mode image of Sample 1 is given in Fig. 10. In both the logarithmic (Fig. 10B) and linear (Fig. 10C) scaled images, only the boundary of the highest weight % gelatin is clearly distinguishable, with an average width of 31.5mm between the logarithmic and linear scales. Based on comparison of the image of the sample (Fig. 10A), and the A-Mode mapping, no reasonable information is gleaned for the 18% and 3.5% gelatin samples using the A-Mode modality.

Unlike standard A-Mode, the three different tissue phantoms can clearly be distinguished by the EBME technique as seen in Fig. 11. From Fig, 11B, EBME shows the 22.5% gelatin to be ~29mm wide with a bulk modulus of 2.45 GPa. The 18.0% and 3.5% gelatin tissue phantoms are ~52mm and 18mm with averaged effective bulk modulus of 2.17 GPa and 1.31 GPa.

The efficacy of the EBME is supported by bulk modulus values determined using standard evaluation method as 2.82 GPa, 2.52 GPa, and 1.71 GPa.

Density values for the three composites are also clearly distinct and able to be used to characterize the sample as a composite of three distinct materials (Fig. 11C). Averaged density and widths from EBME are 1329.4 $kg \cdot m^{-3}$ and 26mm for 22.5% gelatin, 1182.5 $kg \cdot m^{-3}$ and ~54mm for 18.0% gelatin, and 877.3 $kg \cdot m^{-3}$ and 20mm for the lowest ratio 3.5%

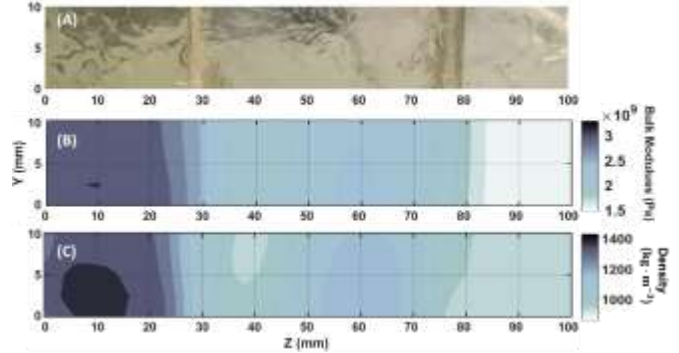

Fig. 11. EBME bulk modulus and density, linearly scaled. (A) is the photographic image of the sample, with materials in the same order as Fig. 11. (B) Effective bulk modulus using Eq. 14 with a color range of 1.5 GPa to 3.25 GPa. (C) Effective density using Eq. 15 where the color bar scales from 900 $kg \cdot m^{-3}$ to 1400 $kg \cdot m^{-3}$.

gelatin tissue phantom.

Ultrasonic images used for evaluation are commonly scaled to a standard material or medium. Here, we created a scaled parameter for the EBME determined bulk modulus and density, where the values are scaled to the ambient medium, water (Fig. 12). The relative bulk modulus, $K_r = K_{sample}/K_{water}$, and the relative density, $\rho_r = \frac{\rho_{sample}}{\rho_{water}}$, serve to indicate the extent of the deviation of an examined material from the ambient medium. All the tissue samples maintain relative values close to that of water. In practical applications, $K_r$ and $\rho_r$ could be calibrated to an ideal sample where $K_r = K_{sample}/K_{Ideal}$ and $\rho_r = \rho_{Sample}/\rho_{Ideal}$, which might be easier to use.

Unlike the standard A-Mode case, the relative values for the samples allow for the 3.5% tissue phantom to be visualized with an apt comparison to water. In medical applications, scaling to healthy tissues values may lead to faster recognition of potentially harmful artifacts in tissue.

For Sample 2, we applied the same techniques to tissue phantoms with compositions that were more similar than Sample 1. Three tissue phantoms of 16.8%, 10.0%, and 6.8% gelatin were synthesized using the method described in ref [35]. The widths of the phantoms were 16mm, 35mm, and 19mm, with densities from standard techniques of 1163 $kg \cdot m^{-3}$, 1106 $kg \cdot m^{-3}$, and 1064 $kg \cdot m^{-3}$, and bulk modulus values of 2.31 GPa, 2.13 GPa, and 2.01 GPa. The low variation in physical properties amongst the phantoms was selected to mimic low variation experienced in tissues in practice.

The reference A-Mode scan of Sample 2 is given in Fig. 13. Both the logarithmic and linear scale figures identify the existence of the lowest concentration tissue phantom as relatively homogenous material on the right (Fig. 13B, 13C).

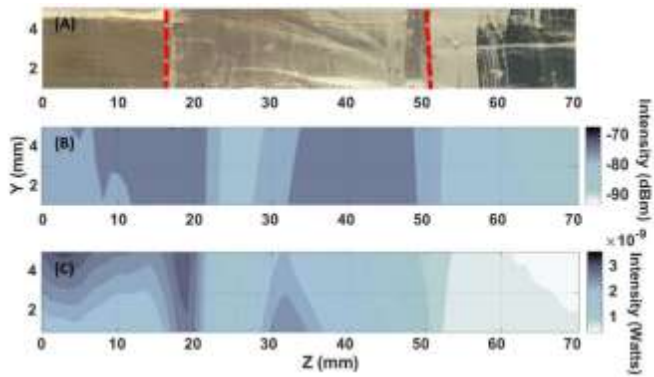

Fig. 13. A-Mode of Sample 2. (A) Image of Sample 2 for reference. Gelatin weights of the tissue phantoms are 16.8% (left – 16mm)), 10.0% (middle-35mm), and 6.8% (right-19mm). (B) Standard A-Mode image scaled logarithmically. (C) Linear scaled A-Mode image of Sample 2.

The relative bulk modulus and density of Sample 2 is shown in Fig. 14 with much greater clarity than standard A-Mode imaging. Visually, $K_r$ most strongly indicates the existence of three distinct materials in the sample (Fig. 14B). The widths of the samples using EBME with scaled elastic values are 18mm for the 16.5% phantom on the left, 30mm for the 10.0% phantom in the center, and 20mm for the lowest gelatin concentration material on the far right of Fig. 14. The estimated widths are 12.5%, -14.3%, and 5.3% off the actual values, but still vastly superior to the A-Mode technique which could not identify three clear materials.

For Sample 2, weak reflection from the tissue phantoms and the lack of application of advanced signal processing techniques led to relatively high SNR as compared with hard materials such as steel. The impact of the low SNR was most strongly manifested in the EBME derived average density and bulk modulus values for the samples which had errors of -7.7%, 8.6%, and 18.4% for the bulk modulus, and 11.2%, -5.6%, and 3.1% for density. Regardless, without advanced signal processing techniques, the EBME method greatly improved the clarity of the samples as compared with standard A-Mode methods while being effective with remote application.

*D. Discussion*

Existing research shows measured mechanical properties and acoustic properties of tissue using ultrasound occupy a range of values instead [38]. Organic tissues normally possess much larger acoustic impedance than water. In this experiment, 6.8% gelatin tissue phantom is representative of very soft liver tissue [34]. The 10% and 16.8% gelatin tissue phantom are representative of tumors at different stages [34, 39]. Prior work on the difference of speed of sound between healthy tissue and tumor tissue has been measured at less than 2% [33]. The density and bulk modulus of tumor tissue, however, has been found to be discernably larger than healthy tissue [39]. The Effective Bulk Modulus elastography (EBME) technique for the remotely determining the bulk modulus and density map in this work was able to clearly distinguish between tissue phantom equivalents of tumorous and healthy tissue. For medical applications, EBME may provide a new path to practical biomedical elastography and tomography applications.

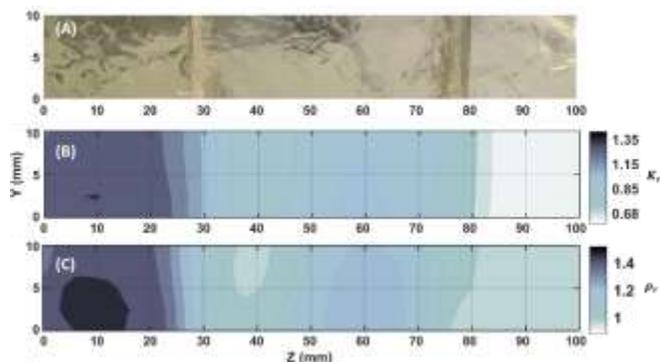

Fig. 12. Relative bulk modulus and density of the Sample 1 as compared to water. (A) Image of the entire scanned area for reference. (B) Relative EBME derived bulk modulus ratio where $K_r = K_{Sample}/K_{water}$. (C) Relative density of the tissue phantoms defined as $\rho_r = \frac{\rho_{sample}}{\rho_{water}}$.

The physical properties of healthy and tumor tissues vary amongst and between states of health of patients [33]. Additionally, the accurate determination of elastic properties using a monostatic setup requires adequate pressure wave reflection which presents challenges for some liquids in soft materials. Elastography maps that use absolute values for the bulk modulus, density, and any other elastic properties may not be the most effective modality for examination of a material. Relative density ($\rho_r$) and relative bulk modulus ($K_r$) can potentially provide greater insight into the examination of a material. Setting the scale according to a fixed density or bulk modulus that is preferred, the user of EBME can qualitatively determine the elastic properties of a sample as shown in Fig. 11. For medical applications, this may be interpreted as setting the relative scales to the values of a known healthy tissue in an analysis, giving the trained eye more efficient guidance to areas of a sample that may be of concern.

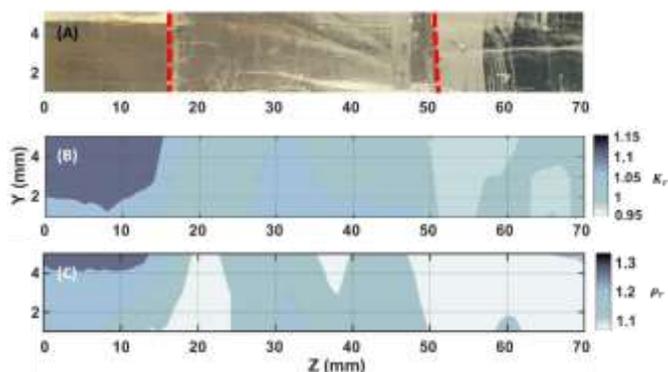

Fig. 14. Relative bulk modulus and density mapping of Sample 2 using EBME. (A) Image of Sample 2 for reference. Gelatin weights of the tissue phantoms are 16.8% (left – 18mm)), 10.0% (middle-33mm), and 6.8% (right-19mm). (B) Relative bulk modulus scaled to water, $K_r = K_{Sample}/K_{water}$. (C) Relative density scaled to water, $\rho_r = \rho_{Sample}/\rho_{water}$.

The elastography technique presented in this study is unique as compared with other existing elastography methods. Table 1 is a comparison of the existing ultrasonic elastography methods. All existing methods need an external force or sonic pulse induced deformation of the sample, unlike EBME. Deformations of a sample may incur unwanted peripheral effects. As with other methods, some knowledge of at least the ambient materials must be known for EBME to be effective. Standard practice for prior methods is the knowledge or assumption of a reference bulk modulus or density for creating elastography maps. In EBME, direct knowledge of the bulk modulus or density of a reference material is not explicitly required.

The need to deform a medium for gathering elastographic information make characterizations of hard materials more difficult. Metals or hard plastic, specifically, do not easily deform without significant applied pressure due to large elasticity constants. Moreover, the linear elasticity strain range is normally small for metals and alloys [40, 41, 42], and especially for hard plastic [43, 44]. Strain map and Poisson's ratio map might be not capable on those materials, because of non-ultrasonic measurable small linear range strain. Large stress easily exceeds the small linear deformation range. Polymer plastics [45, 46], many alloys [47, 48, 49], even polycrystalline metals [50] are usually non-isotropic materials, which their anisotropy were measured by shear waves. Once the results from shear wave maps become directional. Those shear wave methods are not reliable anymore in imaging. EBME, as demonstrated in the experiments above, maintains an advantage over existing techniques it is effective for both hard and soft materials. The lack of a required external, deforming pressure also means EBME is a form of remote sensing that can readily be transportable.

EBME is the first noninvasive elastography method without any external stress or vibration assistant. The main advantage of our EBME is its noninvasive nature and the capability of applying the technique to both hard and soft materials without issue. For imaging soft biomass (M mode), external stress or vibration is able to provide material deformation that can be measured with ultrasound to calculate elasticity. However, hard materials present a particular challenge for techniques that require deformation that's measurable with ultrasound. The linear elastic deformation with applied pressure is normally too small in hard materials to be measured with ultrasound. The lack of need for external stress means EBME can effectively evaluate elasticity in both hard and soft materials without perturbing the sample under evaluation. Though calibration before a scan with a known material will increase the accuracy of the effective bulk modulus as compared to other means, it is not necessary to determine the contrast in elastic properties. For clinical applications, the mechanical properties of the initial material do not vary significantly enough to require calibration for each in vivo test. Transferring the method to practice would require input from physicians and others in the medical field on, at minimum, the primary material encountered in their field.

As mentioned throughout the work, additional signal processing to improve the SNR of the collected data may improve accuracy of EBME method as applied in this work. Additionally, analysis was undertaken with the premise of exclusive use of longitudinal waves as the ambient medium was liquid. Further considerations of transverse waves and their impact on the efficacy of EBME may improve effectiveness of the method as applied to hard and soft materials. This study presented a novel M mode imaging technique for both biomass and industrial materials under controlled, laboratory conditions. Further investigation is needed for the method for addressing complex geometries, advanced signal processing, and automated detection of internal interfaces.

### IV. CONCLUSON

In this study, a new remote, non-destructive monostatic method to produce the bulk modulus elastography and effective density mapping was introduced. The Elastic Bulk Modulus Elastography (EBME) method does not require an external, forced deformation of the material being analyzed, and is functional for both hard and soft materials. The technique does not require explicit knowledge of the elastic properties of a reference material for application. EBME proved effective in discerning between tissue phantoms with small variations in their bulk elastic properties in addition to accurately determining the density and bulk modulus of hard materials. Use of the relative density and bulk modulus of a material using EBME may enable faster detection of unwanted defects in a material. Errors in EBME in this work were primarily a function of low signal to noise ratios from impedance matched tissue phantoms that have low reflectivity in water ambient. In all, the method provides a promising alternative to existing methods for the ultrasonic characterization of the elastic properties of hard and soft materials.

TABLE I
ULTRASONIC ELASTOGRAPHY METHODS COMPARISON

| Methods | Wave Mode | External force | Elasticity values | Water ambient | Input Values |
|---|---|---|---|---|---|
| Strain Map [14, 15, 17] | L | ✓ | ✗ | ✗ | ✗ |
| Impulse strain map [18, 19] | L or T | ✓ | $E$ | ✗ | $Z, f_\sigma, \sigma$ |
| Poisson' ratio map [13, 20] | L | ✓ | $\upsilon$ | ✓ | ✗ |
| Transient shear wave map [29] | T | ✓ | $E$ | ✗ | $\rho, \sigma, f_\sigma$ |
| Point shear wave map [23, 24, 25] | T | ✓ | $E$ or $G$ | ✗ | $\rho, \sigma, f_\sigma$ |
| Surface shear wave map [22] | T | ✓ | $E$ or $G$ | ✗ | $\rho, \sigma, f_\sigma$ |
| **EBME Bulk Modulus** | **L** | **✗** | **$K$** | **✓** | **d, $Z_0$** |
| **EBME Density** | **L** | **✗** | **$\rho$** | **✓** | **d, $Z_0$** |

Table. 1 Comparison between existing ultrasonic elastography techniques and the new, EBME method (**Bold font**) presented in this study. L is the longitudinal mode wave, T is the transverse (shear) mode wave, $E$ is Young's Modulus, $G$ is shear Modulus, $K$ is Bulk Modulus, $\upsilon$ is Poisson's ratio, $\rho$ is sample density, $Z$ is sample material acoustic impedance, $f_\sigma$ is dynamics force frequency, $\sigma$ is applied external stress or radiational force on imaged sample, and $d$ is sample thickness. $Z_0$ is acoustic impedance of reference ambient material. ✓ indicated required and ✗ means not required or none.